\documentclass{article}
\usepackage{amsfonts,amssymb, amsmath,mathrsfs}
\usepackage[english]{babel}

\def\nn{\nonumber }
\def\bq{ \begin{equation} }
\def\eq{ \end{equation} }
\def\ben{ \begin{eqnarray} }
\def\en{ \end{eqnarray} }

\newtheorem{prop}{Proposition}

\newtheorem{re}{Remark}

\newtheorem{exa}{Example}
\newenvironment{exam}{\begin{exa} \rm}{\end{exa}}

\begin{document}

\title{Simple non-Hamiltonian systems with an invariant measure}
\author{A.V. Tsiganov \\
\it\small
St.Petersburg State University, St.Petersburg, Russia\\
\it\small e--mail: andrey.tsiganov@gmail.com}

\date{}
\maketitle
\begin{abstract}
We propose a simple  construction of the non-Hamiltonian dynamical systems  possessing an invariant measure. These non-Hamiltonian systems are  deformations of the Hamiltonian systems associated with  trivial deformations of the canonical Poisson brackets and time transformation.  Sometimes these non-Hamiltonian systems can be identified with known nonholonomic systems and, therefore,  the results are illustrated through the Chaplygin ball and  the Veselova system.
 \end{abstract}

\section{Introduction}
\setcounter{equation}{0}
We consider simple non-Hamiltonian integrable systems which can be obtained from  Hamiltonian systems by a proper change of variables, including time. The main  examples are one of the most famous solvable problems in nonholonomic mechanics, describing rolling of a balanced ball  over a horizontal surface without slipping, is referred to as the Chaplygin ball \cite{ch03} and the nonholomic Veselova system \cite{ves86}.

Let us start with a system of Hamiltonian equations of motion
\[
\dfrac{d}{dt}\,z_k=Z_k\equiv\{H,z_k\}\,,\qquad k=1,\ldots,m\,,
\]
where $\{.,.\}$ being canonical Poisson brackets.  If  $g(z)$ is a nowhere vanishing smooth integrable function, the time reparameterization of the form
\[dt\to g(z)dt\]
 changes the invariant probability measure, despite the fact that the orbits remain invariant and the ergodicity is preserved. Physically, this reflects the fact that the transformed system evolves at different speeds and hence with different residence times along the orbits.

 After time reparameterization one gets the desired simple non-Hamiltonian system possessing an invariant measure.
Associated with the new time vector field
 \[\hat{Z} = g(z)Z\]
is a so-called conformally Hamiltonian vector field. If the initial Hamiltonian vector field $Z$ is complete, then the conformally Hamiltonian vector field $\hat Z$ is a Hamiltonian with respect to another Poisson bracket $\{.,.\}'$. If another technical condition is satisfied, then there is symplectic diffeomorphism which maps each orbit to itself and is an equivariant with respect to the flows of the vector fields $\hat{Z}$ and $Z$ \cite{mar10}.

In classical mechanics transformations $Z\to \hat{Z}$ and $dt\to gdt$ are very useful to investigate the trajectory equivalent systems, see examples in \cite{mar10,ts00,ts01,ves87}.  In nonholonomic mechanics
  we have to  add only one intermediate step to this standard scheme and  to deform  canonical Poisson bracket before the time reparametrization.  This modification plays a key role in reducing non-Hamiltonian systems to Hamiltonian ones \cite{bm01,bm08, bkm11,ch03,jov09,fed89,fed09,bl11}.

 If we have some Hamiltonian vector fields $Z^{(1)},\ldots, Z^{(m)}$, we can consider some more complicated time reparametrization associated with the following vector field
 \[\widetilde{Z}=g_1(z)Z^{(1)}+\cdots+g_m(z)Z^{(m)}.\]
 Such vector fields are useful for Hamiltonization of some nonholonomic systems \cite{ts12rd}.

 Below we prove that transformation of the non-Hamiltonian vector fields $\hat{Z}\to (Z_1,Z_2)\to\widetilde{Z}$ relates the Veselova system with a simple Hamiltonian system on the sphere and then with the nonholonomic Chaplygin ball. It allows us to prove that nonholonomic Veselova and Chaplygin systems are trajectory equivalent systems.

\subsection{Simple non-Hamiltonian systems on cotangent bundles}
Let $Q$ be an $n$-dimensional smooth manifold. Its cotangent bundle $T^*Q $ is naturally endowed with the Liouville $1$-form $\theta$ and symplectic $2$-form $\Omega= d\theta$, whose associated Poisson bivector will be denoted with $P$. In local symplectic coordinates on $T^\ast Q$
\[z=(q,p)=(q_1,\dots,q_n,p_1,\dots,p_n)\]
they have the following local expressions
 \bq\label{poi-0}
 \theta=p_1dq_1+\ldots p_ndq_n\,,\qquad
 \Omega=\sum_{i=1}^n dq_i\wedge dp_i\,,\
 \quad\mbox{and}\quad
 P=\sum_{i=1}^n \dfrac{d}{dq_i}\wedge \dfrac{d}{dp_i}\,,
 \eq
 The $2n$ volume form
\[\mu=\Omega^n\simeq dq\, dp\,,\]
is invariant with respect to Hamiltonian evolution defined by the canonical Poisson bracket
 $\{.,.\}$, the Hamilton function $H$ and the corresponding time $t$
\bq\label{ndef-eqm}
\dfrac{d}{d t}\,q_k=\{H,q_k\}\,,\qquad \dfrac{d}{d t}\,p_k=\{H,p_k\},\qquad k=1,\ldots,n.
\eq
Let us change the form of these Hamiltonian equations (\ref{ndef-eqm}) using trivial deformation of the canonical Poisson bracket. Namely, let us substitute scaling momenta
\bq\label{p-trans}
p_k\to g(q)\,p_k\,\qquad k=1,\ldots,n,
\eq
into the Liouville and symplectic forms
 \[\theta\to \theta_g=g(q)\Bigl(p_1dq_1+\ldots p_ndq_n\Bigr)\,,\qquad \Omega\to \Omega_g=d\theta_g\,.
 \]
 The corresponding Poisson bracket
\bq\label{g-br}
\{q_i,q_j\}_g = 0\,,\qquad \{q_i,p_j\}_g=g(q)\delta_{ij}\,,\quad
\{p_i,p_j\}_g=F_{ij}\,,
\eq
where
\[F_{ij}=\partial_j g\,p_i-\partial_i g\,p_j\,,\qquad
\partial_k=\partial/\partial q_k\,,
\]
is a trivial deformation of canonical Poisson bracket  in the Poisson-Lichnerowicz cohomology \cite{lih77}.
The corresponding volume form
\bq\label{mug}
\mu_g=\Omega_g^{n}\simeq g^{-n}\mathrm dq\, \mathrm dp
\eq
is invariant with respect to a Hamiltonian flow associated with time variable $t_g$ defined by
\bq\label{hamg-eqm}
\dfrac{ d}{ d t_g}\,z=\{H,z\}_g\,.
\eq
Here and below we suppose that $g(q)$ is a nowhere vanishing smooth function on $Q$.
 We consider these equations as an intermediate deformation of the initial Hamiltonian equations of motion (\ref{ndef-eqm}).

\begin{prop}
Transformation of momenta $p\to g(q)\,p$ followed by a change of time
\bq\label{def-time}
d t_d=g(q)\,d t_g
\eq
preserves initial equations of motion (\ref{ndef-eqm}) on $q_k$ and shifts equations on $p_k$
\bq\label{nham-eqm}
\dfrac{d q_k}{d t_d}=\dfrac{d q_k}{d t}\,,\qquad\qquad \dfrac{d p_k}{d t_d}=\dfrac{d p_k}{d t}+\lambda_k q_k\,,\qquad k=1,\ldots,n,
\eq
where
\bq\label{g-q}
\lambda_k=\dfrac{1}{q_k\,g}\,\sum_{j=1}^n F_{jk}\,\dfrac{\partial H}{\partial p_j}\,.
\eq
These non-Hamiltonian equations of motion have invariant volume form
\bq\label{mud}
 \mu_d=g^{1-n}\mathrm dq\, \mathrm dp\,,
\eq
with measure depending on a nowhere vanishing smooth function $g(q)$.
\end{prop}
Proof: The proof is a straightforward verification of the first half of equations
\[
\dfrac{d}{d t_d}\,q_k=g^{-1}\{H,q_k\}_g=g^{-1}\sum_{j=1}^n \{p_j,q_k\}_g\,\dfrac{\partial H}{\partial p_j}\,=\{H,q_k\}=\dfrac{d}{dt}\,q_k\,,
\]
and calculation of the remaining part of equations
\ben
\dfrac{d}{d t_d}\,p_k&=&g^{-1}\{H,p_k\}_g=g^{-1}
\sum_{j=1}^n \left(\{q_j,p_k\}_g\,\dfrac{\partial H}{\partial q_j}
+ \{p_j,p_k\}_g\,\dfrac{\partial H}{\partial p_j}
\right)\nn\\
&=&\dfrac{d}{dt}\,p_k+
g^{-1}\sum_{j=1}^n \{p_j,p_k\}_g\,\dfrac{\partial H}{\partial p_j}=\dfrac{d p_k}{d t}+\lambda_k q_k\,.
\nn
\en
The Jacoby last multiplier $g^{1-n}$ (\ref{mud}) of equations of motion (\ref{nham-eqm}) satisfies to the standard equation
\[
\sum_{j=1}^{2n} \dfrac{\partial}{\partial z_j}\, g^{1-n}\,\dfrac{d z_j}{d t_d}=
\sum_{j=1}^{2n} \dfrac{\partial}{\partial z_j}\, g^{-n}\,\dfrac{d z_j}{d t_g}=0,
\]
because $g^{-n}$ is an invariant measure (\ref{mug}) of the Hamiltonian equations of motion (\ref{hamg-eqm}) . $\square$

Functions $\lambda_i(z)$ in (\ref{nham-eqm}) may be considered as undetermined Lagrange multipliers of some nonholonomic system \cite{bm08}. On the other hand
flow (\ref{nham-eqm}) may be associated with the magnetic geodesic flow on the Riemannian manifold \cite{bol08}.

\begin{exam}
Equations of motion for nonholonomic LR-systems \cite{ves88} are deformations of the standard
Euler equations on the Lie algebra $\mathfrak g$
\bq\label{gves-eqm}
\dfrac{d}{dt}N^i=[\omega,N^i]\,,\qquad \dfrac{d}{dt}M=[\omega,M]+\lambda_iN^i\,,
\eq
in which we shift the second part of initial Euler equations on $\lambda_iN^i$ similar to (\ref{nham-eqm}). Here $M\in \mathfrak g^*$, $\omega\in \mathfrak g$, $[.,.]$ means a coadjoint action and functions $\lambda_i(M,N^i)$ are undetermined Lagrange multipliers. If the singular orbit of  $\mathfrak g$ is symplectomorphic to some  cotangent bundle $T^*Q$ we can try to identify equations (\ref{gves-eqm}) with (\ref{nham-eqm}).
\end{exam}

Natural generalization of the transformation (\ref{p-trans}) looks like
\bq\label{p-transgen}
p_k\to \sum_{i=1}^n L_{ki}(q)\,p_i\,\qquad k=1,\ldots,n,
\eq
so that
\bq\label{tur-def}
\{q_i,q_j\}_L = 0\,,\qquad \{q_i,p_j\}_L=L_{ij}\,,\qquad
\{p_i,p_j\}_L=\sum_{k=1}^n\left(\dfrac{\partial L_{ki}}{\partial q_j}-\dfrac{\partial L_{kj}}{\partial q_i}\right)p_k\,.
\eq
Here $\{.,.\}_L$ is a trivial deformation of the initial Poisson bracket associated with a (1,1) torsionless tensor field $L(q)$, see \cite{Turiel1}. The generic change of time variable $t_g\to t_d$ (\ref{def-time})
\bq\label{def-timegen}
\dfrac{d z}{d t_d}=g_1\{H_1,z\}_{L}+g_2\{H_2,z\}_{L}+\cdots+g_n\{H_n,z\}_{L}\,,
\eq
depends on functions $g_k(z)$ and integrals of motion $H_1,\ldots,H_n$ of the initial Hamiltonian system of equations (\ref{ndef-eqm}).

The Turiel deformation (\ref{p-transgen}) of the initial Poisson bracket yields the following intermediate deformation of the invariant volume form
\[\mu\to \mu_g=\det L(q)\, \mathrm dq\,\mathrm dp\,,\]
whereas transformation of time (\ref{def-timegen}) leads to the next change of this form $\mu_g\to\mu_d$ depending on integrals of motion $H_i$ and functions $g_k(z)$.

The corresponding example of nonholonomic systems describing a rolling of dynamically asymmetric and balanced ball over an absolutely rough fixed sphere may be found in \cite{ts12rd}.

\subsection{Momentum map }
If we suppose that some Lie group $G$ acts on $\mathcal M=T^\ast Q$ via symplectomorphisms, we can try to transfer non-Hamiltonian equations of motion (\ref{nham-eqm}) to the deformed Euler equations (\ref{gves-eqm}) on the Lie algebra $\mathfrak g^*$ using symplectic reduction and momentum map theory \cite{abmar}.

For instance, let $T^{*}\mathbb S^{n-1}\ $ be the cotangent bundle of the sphere $\mathbb S^{n-1}\subset {\mathbb R}^{n}$ realized as the subbundle of $\mathbb S^{n-1}\times {\mathbb R}^{n}$,
\[
T^{*}\mathbb S^{n-1}=\left\{ x,p \in \mathbb S^{n-1}\times \mathbb R^{n};\qquad (p,x) =0\right\}\,,
\]
where $(.,.)$ means the standard scalar product in $\mathbb R^n$. Following to \cite{abmar} let us consider the standard momentum map
\[\varphi:\quad(p,x)\in T^*\mathbb S^{n-1} \to (L,x)\in e^*(n)=so(n)\ltimes \mathbb R^n\,,\qquad
\]
defined by
\bq\label{l-en}
L_{ij}=x_ip_j-x_jp_i\,.
\eq
This standard  mapping relates  cotangent bundle $T^*\mathbb S^{n-1}$ with a singular orbit of the Lie algebra $e^*(n)=so(n)\ltimes \mathbb R^n$. Remind that $e(n)$ is an algebra of the Euclidian group $E(n)$ of motions of the $n$-dimensional Euclidean space $\mathbb R^n$.

It is well known that mapping (\ref{l-en}) is a Poisson map, which transforms canonical Poisson bracket (\ref{poi-0}) on $T^{*}\mathbb S^{n-1}\ $ to the Lie-Poisson bracket on $e^*(n)$:
\ben
\{L_{ij},L_{kl}\}&=&\delta_{kj}L_{il}-\delta_{il}L_{kj} + \delta_{jl}L_{ki} -\delta_{ki}L_{jl}\,,\nn\\
\label{poi-en}\\
\{L_{ij},x_k\}&=&\delta_{ik}x_j-\delta_{jk}x_i\,,\qquad \{x_i,x_k\}=0\,.\nn
\en
It is easy to prove that mapping (\ref{l-en}) is a Poisson map for the deformations of the canonical Poisson bracket $\{.,.\}_g$ (\ref{g-br}) or $\{.,.\}_{L}$ (\ref{tur-def}) on cotangent bundle of the sphere $T^*\mathbb S^{n-1}$, too.

\begin{prop}
Momentum map (\ref{l-en}) transforms the Poisson bracket $\{.,.\}_g$ (\ref{g-br}) on $T^*\mathbb S^{n-1}$ into the Poisson bracket
\ben
\{M_{ij},M_{kl}\}_g&=&g\Bigl(\delta_{kj}M_{il}-\delta_{il}M_{kj} + \delta_{jl}M_{ki} -\delta_{ki}M_{jl}\Bigr)\nn\\
&+&M_{kl}\sum_{m} (\delta_{im}x_j-\delta_{jm}x_i)\partial_m g-M_{ij}\sum_m( \delta_{km}x_l-\delta_{lm}x_k)\partial_m g\,,\nn\\
\label{poi-eng}
\{M_{ij},x_k\}_g&=&g\Bigl(\delta_{ik}x_j-\delta_{jk}x_i\Bigr)\,,\qquad \{x_i,x_k\}_g=0\,.
\en
on the whole space $e^*(n)$. As above here we put $\partial _m=\partial/\partial x_m$.
\end{prop}
Proof is a straightforward verification of all the Poisson bracket properties. $\square$

On a singular orbit of $e^*(n)$  the Poisson brackets (\ref{poi-en}) and (\ref{poi-eng}) are related by a trivial change of variables
\bq\label{trans-em}
 M_{ij}= g(x)L_{ij}\,,\qquad L_{ij}=g^{-1}(x)M_{ij}\,.
\eq
We suppose that on  generic orbit these brackets are related by more complicated transformation discussed in the next paragraph.

\begin{exam} At $n=3$ we can identify an element of $so(3)$ with a vector of the angular momentum
 using well known isomorphism of Lie algebras
 $(so(3),[.,.])$ and $(\mathbb R^3,\times)$:
\[
L_i=\varepsilon_{ijk}L_{jk}\,.
\]
Here $\varepsilon_{ijk}$ is a totally skew-symmetric tensor, $[a,b]$ means commutator in $so(3)$ and $a\times b$ is a vector cross product in $\mathbb R^3$.

In this vector notation the canonical Lie-Poisson bracket (\ref{poi-en}) reads as
\begin{equation}\label{e3}
\bigl\{L_i\,,L_j\,\bigr\}=\varepsilon_{ijk}L_k\,,
 \qquad
\bigl\{L_i\,,x_j\,\bigr\}=\varepsilon_{ijk}x_k \,,
\qquad
\bigl\{x_i\,,x_j\,\bigr\}=0\,,
\end{equation}
whereas the deformed Poisson bracket (\ref{poi-eng}) has the form
\ben\label{pg-e3b}
&&\bigl\{M_i\,,M_j\,\bigr\}_g=\varepsilon_{ijk}\,\left(g(x)M_k
+x_k\sum_{m=1}^3M_m\partial_m g(x)\right)
\,,\\
&&
\bigl\{M_i\,,x_j\,\bigr\}_g=\varepsilon_{ijk}\,g(x)\,x_k \,,
\qquad
\bigl\{x_i\,,x_j\,\bigr\}_g=0\,.
 \nn
\en
These brackets have the following Casimir elements $(x,x)$, $(x,L)$ and $(x,M)$, respectively.

Momentum map (\ref{l-en}) is a symplectomorphism of $T^*\mathbb S^2$ and a singular orbit of $e^*(3)$ defined by
\[(x,L)=0\,.\]
On this singular orbit transformation
\bq\label{lm-e30}
 L_{i}= g^{-1}(x)M_{i}
 \eq
 relates the deformed Poisson brackets (\ref{pg-e3b}) with canonical one (\ref{e3}). In the next paragraph we will consider similar reduction of the Poisson brackets on the generic orbits at $(x,L)\neq 0$.
\end{exam}

\subsection{Simple non-Hamiltonian systems on semi-direct sum of Lie algebras}
Let $\mathfrak g$ be a semisimple Lie algebra, $\mathfrak g^*$ its dual space and
$\mathfrak g=\mathfrak f+\mathfrak p$ its Cartan decomposition. According to \cite{rs94}, let us consider an affine Poisson subspace of the Lax matrices
 \bq\label{lax}
 L(\lambda)=a\lambda+\ell+s\lambda^{-1}\,,\qquad \ell\in\mathfrak f^*,\quad s\in\mathfrak p^*,
 \eq
 passing through the $a\in\mathfrak p^*$. The corresponding Poisson structure on this subspace
 is isomorphic to the Lie-Poisson structure on the semidirect product $\mathfrak f^*\ltimes\mathfrak p^*$:
\[
\{\ell_{ij},\ell_{km}\}\in \mathfrak f^*\,,\qquad
\{\ell_{ij},s_k\}\in \mathfrak p^*\,,\qquad
\{s_i,s_j\}=0\,.
\]
The constant $a$ does not affect kinematics, but is quite useful to produce interesting Hamiltonians for which
equations of motion have the form of the Euler equations
\bq\label{ep-eq}
\dfrac{d}{dt}\,s=\{H,s\}=[s,\omega]\,,\qquad \dfrac{d}{dt}\, \ell=\{H,\ell\}=[\ell,\omega]+[s,b]\,,
\eq
where "angular velocity"\, $\omega=\varphi(\ell)$ depends on the choice of the Hamiltonian $H$, whereas variable $s\in\mathfrak p^*$ plays the role of the "Poisson vector"\,.

All the details may be found in the book \cite{rs94}, another construction of the integrable systems on a semidirect products of a semisimple compact Lie algebra and a commutative algebra may be found in \cite{bol04,bol05,ho09}, see also references within.

Using semisimple structure we can suppose that a natural generalization of the transformations (\ref{p-trans},\ref{trans-em}) looks like
\bq\label{l-trans}
\ell_{ij}\to g(s)\ell_{ij}+h_{ij}(s,c)\,,
\eq
where $g(s)$ is a nowhere vanishing smooth function on commutative part $\mathfrak p^*$, and $h_{ij}$ are functions on $\mathfrak p^*$ depending on the Casimir elements $c=(c_1,\ldots,c_m)$. Such transformation allows as to change initial Poisson brackets $\{.,.\}\to \{.,.\}_g$ on a semidirect product of the Lie algebras similar to (\ref{g-br}).

Now we can add transformation of the time (\ref{def-time}) to the transformation of the variables
 (\ref{l-trans}) in order to get non-Hamiltonian deformations of the initial Euler equations of motion (\ref{ep-eq}):
\bq\label{def-eq}
\dfrac{d}{dt_d}\,s=g^{-1}\{H,s\}_g=[s,\omega]\,,\qquad \dfrac{d}{dt_d}\, \ell=g^{-1}\{H,\ell\}_g=[\ell,\omega]+[s,b]+\lambda_ks_k\,.
\eq
Functions $\lambda_k$ may be considered as undetermined Lagrange multipliers associated with some constraints added to the initial equations (\ref{ep-eq}).

\begin{exam}
Let us continue to consider the Lie algebra $e^*(3)$. On its generic orbit the counterpart
of transformation (\ref{p-trans},\ref{lm-e30}) has the following form
 \begin{equation} \label{trans-lm}
L_i=g^{-1}(x)\Bigl( M_i-b\,h_i(x)\Bigr)\,,\qquad b=(x,M)\,,
 \end{equation}
where three functions $h_i(x)$ satisfy to one algebraic equation
\[c\,g(x) = 1+\sum_{i=1}^3 x_ih_(x)\,,\qquad \Leftrightarrow\qquad (x,L)=c\,(x,M)\,,\qquad c\in\mathbb R\,,\]
and one differential equation
\[
\sum_{i=1}^3 \Bigl[
h_i^2-\partial_i h_i+h_i\sum_{j\neq i} (x_j\partial_i h_j-x_i\partial_jh_j)
\Bigr]=0\,,\qquad \partial_k=\dfrac{\partial}{\partial x_k}\,.
\]
In this case mapping (\ref{trans-lm}) reduces deformed Poisson bracket (\ref{pg-e3b}) to canonical Poisson brackets (\ref{e3}) on generic orbits of $e^*(3)$. In Section \ref{sec-ch} we present particular solutions of these equations associated with the Chaplygin nonholonomic ball.
\end{exam}

\section{Integrable non-Hamiltonian systems on the sphere}
\setcounter{equation}{0}
Transformations (\ref{p-trans}) and (\ref{def-time}) change the form of the equations of motion and preserve the main invariant geometric and topological properties of the Hamiltonian system (\ref{hamg-eqm}). For instance, it is clear that a change of time (\ref{def-time}) or (\ref{def-timegen}) preserves integrability. In order to get such integrable Hamiltonian and the corresponding non-Hamiltonian systems we can use the Jacobi separation of variables method \cite{abmar}, method of the classical $r$-matrix \cite{rs94}, the argument shift method \cite{bol04,bol05,ho09} etc.

For instance, in the Jacobi method we can substitute some known variables $u_1,\ldots,u_n$ (elliptic, parabolic, conical, prolate spheroidal or other) and conjugated moments $p_{u_k}$ into the separated relations
\bq\label{sep-rel}
\Phi_i(u_i,p_{u_i},H_1,\ldots,H_n)=0\,,\qquad i=1,\ldots,n\,,\qquad\mbox{so that}\quad\det\left[\frac{\partial \Phi_i}{\partial H_j}\right]
\not=0\,.
\eq
Solving the resulting system of algebraic equations with respect to $H_1,\ldots,H_n$ one gets integrable Hamiltonian system, which may be interesting in application. After that we can get integrable non-Hamiltonian systems (\ref{nham-eqm}) using a proper change of time.

Let us consider few integrable Hamiltonian systems (\ref{ndef-eqm}) and non-Hamiltonian systems (\ref{nham-eqm} ) separable in an orthogonal coordinate system on a sphere $\mathbb S^{n-1}$. Remind, that
 all orthogonal separable coordinate systems can be viewed as an orthogonal
sum of certain basic coordinate systems \cite{kal}. All these basic systems on $\mathbb S^{n-1}$ may be obtained from the elliptic coordinate system $u=(u_1,\ldots,u_{n-1})$
with parameters $e_1< e_2 <\cdots < e_{n}$ defined through the equation
\bq\label{ell-sph}
e(\lambda)=\sum_{i=1}^{n} \dfrac{x_i^2}{\lambda-e_i}=\dfrac{\prod_{k=1}^{n-1}(\lambda-u_k)}{\prod_{i=1}^{n}(\lambda-e_i)}\,.
\eq
 Here $\sum x_i^2 = 1$. Like the elliptic coordinates in $\mathbb R^n$, the
elliptic coordinates on $\mathbb S^{n-1}$ are also orthogonal and only locally defined. They
take values in the intervals
\[e_1 < u_1 < e_2 < u_2 < \cdots < u_{n -1}< e_{n}\,.\]
The coordinates and the parameters can be subjected to a simultaneous linear
transformation $x_i\to ax_i + b$, $e_i\to ae_i + b$, so it is always possible to choose
$e_1 = 0$ and $e_n = 1$. The coordinates can be degenerated by letting some, but
not all, parameters $e_i$ coincide.

Let us substitute elliptic coordinates $u$ and conjugated momenta $p_u$ into the uniform separated relations of the St\"ackel form
\bq\label{ch-seprel}
f(u_j)\prod_{i=1}^{n}(u_j-e_i)p_{u_j}^2=u_j^{n-2}H_{n-1}+\ldots+u_jH_2+H_1+V(u_j),\qquad j=1,\ldots,n-1\,,
\eq
where rational functions $f(u)$ and $V(u)$ define the St\"ackel metric and the potential \cite{ts99,ts00} , respectively. Solving resulting equations with respect to $H_1,\ldots,H_{n-1}$ one gets the desired tuple of integrals of motion on $T^\ast\mathbb S$ for the Hamiltonian (\ref{ndef-eqm}) and non-Hamiltonian equations of motion (\ref{nham-eqm}).

It is easy to see, that generic separated relations (\ref{ch-seprel}) may be obtained from the separated equations with $f(u)=1$
\bq\label{zch-seprel}
\prod_{i=1}^{n}(u_j-e_i)p_{u_j}^2=u_j^{n-2}H_{n-1}+\ldots+u_jH_2+H_1+V(u_j),\qquad j=1,\ldots,n-1\,.
\eq
using the transformation
 \bq\label{p-trans2}
 p_{u_j}\to \sqrt{f(u_j)\,}p_{u_j}\,,
 \eq
 which changes the so-called St\"ackel matrix and the corresponding metric \cite{ts00,ts01}.
 Below we will combine this transformation (\ref{p-trans2}) acting on the form of the separated relations with the similar transformation (\ref{p-trans}) changing the symplectic structure and with the time reparametrization (\ref{def-time}).

\begin{exam}
Separated relations (\ref{zch-seprel}) with $f(u)=1$ yield many well-known integrable two-dimensional Hamiltonian systems on a sphere including the Euler top, Neumann system, Braden and Ro\-so\-cha\-ti\-us systems. For instance, if $n$=2 and $V=0$, the solutions $H_{1,2}$ of the corresponding separated equations
\[
\varphi_3(u_j)\,p_{u_j}^2=-u_jH_2+H_1\,,\qquad \varphi_3(u)=4\prod_{i=1}^{3}(u-e_i)\,,\qquad j=1,2,
\]
coincide with the Euler integrals of motion
 \[H_1=(L,\mathbf AL)\,,\qquad H_2=(L,L)\,,\qquad \mathbf A=diag(e_1,e_2,e_3)\,,\]
where $L$ is the angular momentum vector with the entries
\bq\label{l-ell}
L_i= -\dfrac{2x_jx_k(e_j-e_k)}{u_1-u_2}\Bigr((e_i-u_1)p_{u_1}-(e_i-u_2)p_{u_2}\Bigl)\,,\qquad i=1,2,3\,,
\eq
At arbitrary $n$ and $V$ description of the corresponding integrable systems on $\mathbb S^n$ may be found  in \cite{bog86,rw85}.

\end{exam}

\begin{exam}
 If we combine the transformations (\ref{p-trans}), (\ref{def-time}) and (\ref{p-trans2}) at
\[
f(u)=\dfrac{u+1}{(e_1+1)\cdots(e_n+1)}\,,\qquad
g=\sqrt{-e(-1)}=\sqrt{\dfrac{f(u_1)\ldots f(u_{n-1})}{(e_1+1)\cdots(e_n+1)}}\,,
\]
one gets the $n$-dimensional generalization of the nonholonomic Chaplygin ball. Here $e(\lambda)$ is a rational function incoming into the definition (\ref{ell-sph}) of the elliptic coordinates on the sphere.

For instance, let us put $n=2$ and $V=0$. In this case the solutions $H_{1,2}$ of the corresponding separated relations
\[
\dfrac{(u_j+1)}{(e_1+1)(e_2+1)(e_3+1)}\,\varphi_3(u_j)\,p_{u_j}^2=-u_j(H_2-dH_1)+dH_1\,,\qquad j=1,2,
\]
coincide with the integrals of motion for the nonholonomic Chaplygin ball (\ref{3-int})
\[H_1=(M,\mathbf A_gM)\,,\qquad H_2=(M,M)\,,\]
if  $e_j=dI_j^{-1}$ following to \cite{ch03}. Here $I_j$ are elements of the inertia tensor, $d$ is some parameter and $M$ is an angular momentum
 \[
 M_i= -\dfrac{2g(x)\,x_jx_k(e_j-e_k)}{u_1-u_2}\Bigr((e_i-u_1)p_{u_1}-(e_i-u_2)p_{u_2}\Bigl),
 \]
so that $(x,M)=0$. All the details will be presented in Section \ref{sec-ch}.

 Nonholonomic systems on the sphere with $V\neq 0$ and the corresponding Lax matrices are considered in \cite{ts11ch,ts12rd}. Discussion of the $n$-dimensional generalizations of the Chaplygin ball may be found in \cite{jov09,ho09}. Note that the restrictions on parameters $e_j$ in \cite{jov09,ho09} are related with a special choice of the nonholonomic constraint.
\end{exam}

Using other functions $f(u)$ (\ref{p-trans2}) , $V(u)$ (\ref{zch-seprel}) and $g(u)$ (\ref{p-trans},\ref{def-time}) we can get a lot of non-Hamiltonian integrable systems on the sphere. For instance, if $f=u$ and $g=\sqrt{f_1\cdots f_n}$, then we get the so-called Veselova system and its generalizations, see Section \ref{sec-ves}. The main problem is how to find non-Hamiltonian systems, which are really applicable in mechanics.

\section{Nonholonomic Chaplygin ball}\label{sec-ch}
\setcounter{equation}{0}
The nonholonomic Chaplygin ball \cite{ch03} is  a dynamically balanced 3-dimensional ball that rolls on a horizontal table without slipping or sliding. `Dynamically balanced' means that the geometric center coincides with the center of mass, but the mass distribution is not assumed to be homogeneous. Because of the roughness of the table this ball cannot slip, but it can turn about the vertical axis without violating the constraints.

 Let $\omega$ and $v$ be the angular velocity and velocity of the center of mass of the ball expressed in the so-called body frame. Its mass, inertia tensor and radius will be denoted by $m$, $\mathbf I$ and $a$ respectively. This reference frame is firmly attached to the body, its origin is located at the center of mass of the body, and its axes coincide with the principal inertia axes of the body. The inertia tensor of the body in this frame is diagonal $\mathbf I=\mbox{\rm diag}(I_1,I_1,I_3)$.

Since slipping at the contact points is absent, its velocity vanishes and we have the following nonholonomic constraint
\bq\label{rel-ch}
v+\omega\times r=0\,.
\eq
Here $r$ is the vector joining the center of mass with the contact point expressed in the body frame and $\times$ means the vector product in $\mathbb R^3$.

The angular momentum $M$ of the ball with respect to the contact point with the plane is equal to
\bq\label{om-m}
M=(\mathbf I+d\mathbf E)\omega-d( x,\omega) x\,,\qquad \qquad d=ma^2\,.
\eq
Here $x$ is the unit normal vector to the fixed sphere at the contact point so $r=-ax$, $\mathbf E$ is the unit matrix and $(.,.)$ means the standard scalar product in $\mathbb R^3$.

Using constraint (\ref{rel-ch}) we can eliminate the Lagrange multiplier $\lambda$ from the initial equations of motion
\[
m\dot{v}=mv\times \omega+\lambda\,, \qquad \mathbf I\dot{\omega}=\mathbf I \omega\times\omega+\lambda r\,,\qquad \dot{x}=x\times \omega\,,
\]
and obtain the reduced equations of motion in the $x,M$- variables
\bq\label{m-eq}\dot M=M\times \omega\,,\qquad \dot x=x\times \omega\,.
\eq
All the details may be found in \cite{ch03,bm08}.

For further use we rewrite the relation (\ref{om-m}) in the following equivalent form
\bq\label{w12}
 \omega=\mathbf A_g\,M\equiv\Bigl( \mathbf A +{d}\, g^{-2}\,\mathbf A\,\bigl(x \otimes x\bigr)\,\mathbf A \Bigr) M\,,
\eq
where
\bq\label{a-mat}
\mathbf A=(\mathbf I+d\mathbf E)^{-1}\equiv\left(
 \begin{array}{ccc}
 a_1 & 0 & 0 \\
 0 & a_2 & 0 \\
 0 & 0 & a_3
 \end{array}
 \right)\,,\qquad \mathrm g=\sqrt{1-d ( x,\mathbf A x)}\,.
\eq
There are four integrals of motion
\bq\label{3-int}
H_1=(M,\mathbf A_g M)\,,\qquad H_2=(M,M)\,,\qquad H_3=( x, x)\,,\qquad H_4=( x,M)
\eq
and an invariant measure
\bq\label{rho}
\mu_{ch}= g^{-1}\mathrm d x \mathrm dM\,,\qquad g=\sqrt{1-d ( x,\mathbf A x)}\,.
\eq
Thus, the six equations (\ref{m-eq}) are integrable in quadratures by the Euler-Jacobi theorem.

According to \cite{bm01, bm08}, integrals of motion (\ref{3-int}) are in  involution with respect to the Poisson bracket
\bq\label{3-br}\\
\{M_i,M_j\}_g=\varepsilon_{ijk}\left(g{M_k}-\dfrac{d(M, \mathbf A x)}{g} x_k\right),\quad
 \{M_i, x_j\}_g=\varepsilon_{ijk}g x_k\,,\quad
\{ x_i, x_j\}_g=0,
\eq
where $\varepsilon_{ijk}$ is a totally skew-symmetric tensor. This bracket coincides with the Poisson bracket
(\ref{pg-e3b}) obtained from (\ref{p-trans}) by using a standard momentum map.

 Equations of motion (\ref{m-eq}) have the form (\ref{def-eq})
 \begin{equation}\label{new-eq}
\dfrac{dz}{dt}=g^{-1}\{H,z\}_g\,,\qquad H=\dfrac{H_1}{2}\,.
\end{equation}
After changing the time variable (\ref {def-time})
 \begin{equation} \label{time-ch}
dt\to gdt
 \end{equation}
these non-Hamiltonian equations of motion are reduced to the Hamiltonian equations with respect to the deformed Poisson bracket (\ref{3-br}).

We can reduce this trivially deformed Poisson bracket to canonical one.
\begin{prop}
The Poisson map ( \ref{trans-lm})
\begin{equation}\label{trans-lmf}
\begin{array}{l}
L_1=g^{-1}\left(M_1-\dfrac{bx_1}{(x,x)}\left(1+\dfrac{x_3^2}{\nu}\right)\,
\right)+\dfrac{cx_1}{x_1^2+x_2^2}\,,
\\
\\
L_2=g^{-1}\left(M_2-\dfrac{bx_2}{(x,x)}\left(1+\dfrac{x_3^2}{\nu}\right)\,
\right)+\dfrac{cx_2}{x_1^2+x_2^2}\,,
\\
\\
L_3=g^{-1}\left(M_3
-\dfrac{bx_3}{(x,x)}\left(1-\dfrac{x_1^2+x_2^2}{\nu}\right)\,
\right)\,
\end{array}
 \end{equation}
reduces the deformed Poisson bracket (\ref{3-br}) to a canonical Li-Poisson bracket (\ref{e3}) on $e^*(3)$. Here the Casimir elements are equal to
 \[ (x,L)=c\,,\qquad (x,M)=b\,, \]
and
 \[\nu=x_1^2+x_2^2-d(x,x)(a_1x_1^2+a_2x_2^2)\,.\]
 \end{prop}
 The proof is a straightforward calculation of the Poisson brackets (\ref{e3}) . $\square$

Summing up, using the transformation of coordinates including time we can reduce the initial non-Hamiltonian dynamical system (\ref{m-eq}) to a new Hamiltonian dynamical system on $e^*(3)$ with a canonical Poisson bracket. In the next paragraph we will study some properties of this Hamiltonian system.

\subsection{Hamiltonian system on sphere associated with the Chaplygin ball}
The Poisson map
\begin{equation}\label{invtrans-lmf}
\begin{array}{l}
M_1=g\left(L_1-\dfrac{cx_1}{x_1^2+x_2^2}\right)
+\dfrac{bx_1}{(x,x)}\left(1+\dfrac{x_3^2}{\nu}\right)\,,
\\
\\
M_2=g\left(L_2-\dfrac{cx_2}{x_1^2+x_2^2}
\right)+\dfrac{bx_2}{(x,x)}\left(1+\dfrac{x_3^2}{\nu}\right)\,,
\\
\\
M_3=gL_3+
\dfrac{bx_3}{(x,x)}\left(1-\dfrac{x_1^2+x_2^2}{\nu}\right)\,,
\end{array}
 \end{equation}
is an inverse transformation to (\ref{trans-lmf}). Applying this map to initial integrals of motion
 (\ref{3-int}) one gets integrable Hamiltonian system on $e^*(3)$.

Moreover, at $c=0$ one gets a particular Poisson map (\ref{trans-lmf}), which relates generic symplectic leaves of the deformed Poisson bracket $\{.,.\}_g$ (\ref{3-br}) and singular symplectic leaves of canonical Lie-Poisson bracket $\{.,.\}$ (\ref{e3})
\[
(x,L)=0\,,
\]
which are symplectomorphic to the cotangent bundle $T^*\mathbb S^2$ of the sphere.

In this case initial integrals of motion (\ref{3-int}) are reduced to the following integrals of motion on $T^*\mathbb S^2$
\ben
\widetilde{H}_1&=&g^2(L,\mathbf A_g L)+\dfrac{2bg}{\nu}\Bigl((x,\mathbf AL)-a_3x_3L_3\Bigr)+\dfrac{b^2(a_1x_1+a_2x_2)\bigl(1-d(a_1x_1^2+a_2x_2^2)\bigr)}{\nu^2}\nn\\
\label{int-ch0}\\
\widetilde{H}_2&=&g^2(L,L)+\dfrac{2bgx_3L_3}{\nu}+b^2\left(1+\dfrac{x_3^2(x_1^2+x_2^2)}{\nu^2}\right)\,.
\nn
\en
In terms of elliptic coordinates $u_{1,2}$ (\ref{ell-sph}) on the sphere and the corresponding momenta $p_{u_{1,2}}$ (\ref{l-ell}) these integrals looks like
\begin{eqnarray}
\widetilde{H}_1&=&\mathrm ap_{u_1}^2+2\mathrm bp_{u_1}p_{u_2}+\mathrm c p_{u_2}^2+\mathrm d p_{u_1}+\mathrm e p_{u_2}+\mathrm f\,,\nonumber\\
\label{qua-st}\\
\widetilde{H}_2&=&\mathrm Ap_{u_1}^2+2\mathrm Bp_{u_1}p_{u_2}+\mathrm Cp_{u_2}^2+\mathrm Dp_{u_1}+\mathrm Ep_{u_2}+\mathrm F\,.\nonumber
 \end{eqnarray}
 Here $e_i=dI_i^{-1}$ and the coefficients of the polynomials in (\ref{qua-st}) are analytical functions on $u_{1,2}$.

 According to \cite{sok}, the quasi-St\"ackel integrals of motion (\ref{qua-st})
 admit partial separation of variables in coordinates $s_{1,2}$ defined as the roots of the equation
\[
(\mathrm B-\mathrm bs)^2-(\mathrm A-\mathrm as)(\mathrm C-\mathrm cs)=0\,.
\]
 In our case we have
 \[s_{1,2}=d(1+u_{1,2})u^{-1}_{1,2}.\]
 Similar to the Clebsch top \cite{sok}, for the Chaplygin ball these variables evolve on Prym variety of some non-hyperelliptic curve, in contrast with the Chaplygin variables of separation \cite{ch03,bkm11}.

\section{The Veselova system}\label{sec-ves}
\setcounter{equation}{0}
The Veselova system \cite{ves86} describes motion of the rigid body with a fixed point under
constraint
\bq\label{rel-ves}
(\omega,x)=0\,.
\eq
Here $\omega=\hat{\mathbf A} M$ is the vector of the angular velocity in the body frame, $x$ is a unit vector which is fixed in a space frame and $(.,.)$ denotes the standard scalar product in $\mathbb R^3$.

The corresponding equations of motion are equal to
\bq\label{ves-eqm}
\dot{M}=M\times \omega +\lambda x\,,\qquad \dot{x}=x\times \omega\,.
\eq
Here $M$ is the angular momentum vector in the body frame and $\hat{\mathbf A}$ is the diagonal non degenerate matrix
\[
\hat{\mathbf A}=\left(
 \begin{array}{ccc}
 \hat{a}_1 & 0 & 0 \\
 0 & \hat{a}_2 & 0 \\
 0 & 0 & \hat{a}_3 \\
 \end{array}
 \right)\,.
\]
In order to eliminate the undetermined Lagrangian multiplier $\lambda$ we have to differentiate the constraint (\ref{rel-ves}) and
 find $\lambda$ using equations of motion (\ref{ves-eqm})
\[
\lambda=\dfrac{(\hat{\mathbf A} M\times M, \hat{\mathbf A} x)}{(\hat{\mathbf A} x,x)}\,.
\]
The resulting equations of motion have four integrals of motion
\[
\hat{H}_1=(M,\hat{\mathbf A}M)\,,\qquad \hat{H}_2=(M,M)-(x,x)^{-1}(x,M)^2\,,\qquad \hat{H}_3=(x,x),\qquad
\hat{H}_4=(x,\omega)\,,
\]
and an invariant measure
\bq\label{rho-ves}
\mu_{ves}= \hat{g}\,\mathrm d x \mathrm dM\,,\qquad \hat{g}(x)={ ( x,\mathbf A x)}^{1/2}\,.
\eq
If the constraint (\ref{rel-ves}) holds then these integrals of motion are in involution with respect to the following Poisson brackets
\ben
&&\bigl\{M_i\,,M_j\,\bigr\}_{ag}=\dfrac{\varepsilon_{ijk}}{a_ia_j}\,\left(a_k\hat{g}^{-1}M_k
+x_k\sum_{m=1}^3a_mM_m\partial_m \hat{g}^{-1}\right)
\,\nn \\
\nn\\
&&
\bigl\{M_i\,,x_j\,\bigr\}_{ag}=\varepsilon_{ijk}\,\hat{g}^{-1}\,a_i^{-1}\,x_k \,,
\qquad
\bigl\{x_i\,,x_j\,\bigr\}_{ag}=0\,.
\nn
\en
Reduction of the Poisson brackets $\{.,.\}_{ag}$ to canonical Lie-Poisson brackets (\ref{e3}) and equivalence
of the Veselova system to the Chaplygin ball without spinning are discussed in \cite{bm08}.

\section{Generalized Veselova system}
 The Veselova system is an LR system on the Lie group $G = SO(3)$ \cite{fed09, ves88} or on the Lie algebroid \cite{clm09}.

 In \cite{bm08,fed89} "mathematical"\, equations of motion (\ref{ves-eqm}) were rewritten in the "physical"\, form
\[
\mathbf I\dot{\omega}=\mathbf I\omega\times \omega+\lambda x\,,\qquad \dot{x}=x\times \omega\,,
\]
where $\mathbf I=\hat{\mathbf A}^{-1}$ being the inertia tensor of the rigid body and undetermined Lagrangian multiplier
\[
\lambda=-\dfrac{(\mathbf I^{-1}x,\mathbf I\omega\times\omega)}{(\mathbf I^{-1}x,x)}
\]
was obtained from the mechanical reason stating an absence of the spinning around vector $x$.

In this section we continue to study the equations of motion (\ref{ves-eqm}) in their initial form \cite{ves86,ves88}. However, we will consider generalized Veselova system at
 \bq\label{rel-fed}
(\omega,x)=b\,,\qquad b\in\mathbb R\,,
\eq
instead of (\ref{rel-ves}) following to \cite{fed89}. It means that the projection of the angular velocity vector $\omega$ on $x$ is a non-zero constant $b$.

Let us introduce vector $ K$ instead of the angular momentum vector $M$
\[
K=M-\dfrac{\Bigl((\mathbf E-\hat{\mathbf A})M,x\Bigr)}{(x,x)}\,,
\]
so that initial equations of motion (\ref{ves-eqm}) for the Veselova system read as
\bq\label{eqm-fed}
\dot{K}=K\times \omega \,,\qquad \dot{x}=x\times \omega\,,\qquad\mbox{where}\qquad \omega=\hat{\mathbf A} M\,,
\eq
whereas constraint (\ref{rel-fed}) takes the form of
\[
 (x,K)=b\,.
\]
Below we will identify $(x,K)$ with the Casimir element of the Poisson bracket.

It is easy to see that the Jacobian of this transformation of coordinates is proportional to the Jacobi multiplier
\[
\mathbf J=\det\left|\dfrac{\partial (x,M)}{\partial (x,K)}\right|=\hat{g}^2(x)\equiv \hat{a}_1x_1^2+\hat{a}_2x_2^2 +\hat{a}_3x_3^2\,.
\]
So, the transformation of coordinates $M\to K$ is invertible where  invariant measure $\mu_{ves}$ (\ref{rho-ves}) exists and where  change of time exists, which allows us to integrate equations of motion in quadratures \cite{ves86}.

In variables $x,K$ invariant measure of the equations of motion (\ref{eqm-fed}) is expressed via old measure $\mu_{ves}$ and Jacobian
\bq\label{fed-rho}
\mu_v= \mu_{ves}\mathbf J^{-1}= \hat{g}^{-1}\,\mathrm d x \mathrm dK\,,\qquad \hat{g}(x)={ ( x,\hat{\mathbf A} x)}^{1/2}\,.
\eq
It is easy to prove that integrals of motion
\bq\label{int-fed}
\hat{H}_1=(K,\hat{\mathbf A}_g K)\,,\quad
\hat{H}_2=(K,K)\,,\quad \hat{H}_3=(x,x)\,,\quad \hat{H}_4=(x,K)\,,
\eq
where
\[
\hat{\mathbf A}_g=\hat{\mathbf A}-\hat{g}^{-2}\,(\mathbf E-\hat{\mathbf A})\,(x\otimes x)\,(\mathbf E-\hat{\mathbf A})\,,
\]
are in the involution with respect to the Poisson brackets
\ben
&&\bigl\{K_i\,,K_j\,\bigr\}_{v}=\varepsilon_{ijk}\,\hat{g}K_k
+\varepsilon_{ijk}\,x_k\left(\sum_{m=1}^3K_m\partial_m \hat{g}-\dfrac{b}{\hat{g}}\right)
\,,\nn\\
\label{p-fed}\\
&&
\bigl\{K_i\,,x_j\,\bigr\}_{v}=\varepsilon_{ijk}\,\hat{g}\,x_k \,,
\qquad
\bigl\{x_i\,,x_j\,\bigr\}_{v}=0\,.
 \nn
\en
As usual $(x,x)$ and $(x,K)$ are the Casimir functions for these brackets and non-Hamiltonian equations of motion (\ref{eqm-fed})
\bq\label{new-eq-ves}
\dfrac{d}{dt}\,z=\hat{g}^{-1}(x)\left\{H,z\right\}_{v}\,,\qquad H=\dfrac{\hat{H}_1}{2}\,,
\eq
become Hamiltonian equations after the change of time $dt\to \hat{g}dt$.

\begin{prop}
If $(x,x)=1$ and parameters $\hat{a}_i$ of the Veselova system are related with parameters $a_i$ of the Chaplygin system
 \bq\label{tr-par}
 \hat{a}_i=1-da_i\,,
\eq
then the Poisson brackets $\{.,.\}_v$ (\ref{p-fed}) and invariant measure $\mu_v$ (\ref{fed-rho}) coincide with the Poisson brackets $\{.,\}_g$ (\ref{3-br}) and invariant measure $\mu_{ch}$ (\ref{rho}).

 In this case integrals of motion $\hat{H}_{i}$ (\ref{int-fed}) for the Veselova system are expressed via integrals of motion $H_i$ (\ref{3-int}) for the Chaplygin ball
\bq\label{int-ves0}
\hat{H}_1=H_2-dH_1\,,\qquad
\hat{H}_2=H_2\,,\qquad \hat{H}_3=H_3\,,\qquad \hat{H}_4=H_4\,.
\eq
\end{prop}
 The proof is a straightforward. $\square$

Consequently, the Veselova system the Poisson brackets $\{.,.\}_v$ are reduced to canonical Lie-Poisson brackets on $e^*(3)$ using the combination of (\ref{trans-lmf}) and (\ref{tr-par})
\bq\label{trans-e3ves}
\begin{array}{l}
L_1= \hat{g}^{-1}\left(K_1-\dfrac{bx_1}{(x,x)}\left(1+\dfrac{x_3^2}{\nu}\right)\right)+\dfrac{cx_1}{x_1^2+x_2^2}\,,\\
\\
L_2= \hat{g}^{-1}\left(K_2-\dfrac{bx_2}{(x,x)}\left(1+\dfrac{x_3^2}{\nu}\right)\right)+\dfrac{cx_2}{x_1^2+x_2^2}\,,\\
\\
L_3= \hat{g}^{-1}\left(K_3-\dfrac{bx_3}{(x,x)}\left(1-\dfrac{x_1^2+x_2^2}{\nu}\right)\right)\,,\qquad
\nu=\hat{a}_1x_1^2+\hat{a}_2x_2^2\,.
\end{array}
\eq
Inverse mapping is the combination of (\ref{invtrans-lmf}) and (\ref{tr-par}) as well.

If we put $c=0$, the combination of transformations  (\ref{new-eq-ves}) and  (\ref{trans-e3ves}) relates the nonholonomic Veselova system with integrable Hamiltonian system on the sphere (\ref{int-ch0}) . The corresponding integrals of motion look like
\bq
\hat{H}_1=\widetilde{H}_2-d\widetilde{H}_1\,,\qquad
\hat{H}_2=\widetilde{H}_2\,.
\eq
So, nonholonomic Veselova system is a trajectory equivalent to the Hamiltonian system on the sphere (\ref{int-ch0}) and, therefore, to the nonholonomic Chaplygin ball. These three dynamical systems have a common Lagrangian foliation and the common trajectories considered us unparametrized curves.

Note that this equivalence explains the applicability of the Chaplygin transformation \cite{ch03} to the Veselova system \cite{fed89}.

\end{document}